\documentclass[manuscript,screen,nonacm]{acmart}

\setcopyright{none}
\AtBeginDocument{%
  }

\acmISBN{978-1-4503-XXXX-X/2018/06}

\usepackage{booktabs}
\usepackage{array}
\usepackage{adjustbox}
\usepackage[table]{xcolor}
\usepackage{multirow}
\usepackage{graphicx}
\usepackage{placeins}
\usepackage{float}

\begin{document}

\title{Ambiguity Collapse by LLMs: A Taxonomy of Epistemic Risks}

\author{Shira Gur-Arieh}
\affiliation{
  \institution{Harvard University}
  \city{Cambridge}
   \country{United States}
}
\email{sgurarieh@sjd.law.harvard.edu}

\author{Angelina Wang}
\affiliation{
  \institution{Cornell Tech}
  \city{New York City}
   \country{United States}
}
\email{angelina.wang@cornell.edu}

\author{Sina Fazelpour}
\affiliation{
  \institution{Northeastern University}
  \city{Boston}
   \country{United States}
}
\email{s.fazel-pour@northeastern.edu}

\renewcommand{\shortauthors}{Gur-Arieh et al.}

\begin{abstract}
  Large language models (LLMs) are increasingly used to make sense of ambiguous, open-textured, value-laden terms. Platforms routinely rely on LLMs for content moderation, asking them to label text based on disputed concepts like “hate speech” or “incitement”; hiring managers may use LLMs to rank who counts as “qualified”; and AI labs increasingly train models to self-regulate under constitutional-style ambiguous principles such as “biased” or “legitimate.” This paper introduces \emph{ambiguity collapse}: a phenomenon that occurs when an LLM encounters a term that genuinely admits multiple legitimate interpretations, yet produces a singular resolution, in ways that bypass the human practices through which meaning is ordinarily negotiated, contested, and justified. Drawing on interdisciplinary accounts of ambiguity as a productive epistemic resource, we develop a taxonomy of the epistemic risks posed by ambiguity collapse at three levels: process (foreclosing opportunities to deliberate, develop cognitive skills, and shape contested terms), output (distorting the concepts and reasons agents act upon), and ecosystem (reshaping shared vocabularies, interpretive norms, and how concepts evolve over time). We illustrate these risks through three case studies, and conclude by sketching multi-layer mitigation principles spanning training, institutional deployment design, interface affordances, and the management of underspecified prompts, with the goal of designing systems that surface, preserve, and responsibly govern ambiguity.
\end{abstract}




\maketitle

\section{Introduction}\label{sec:introduction}
Large language models (LLMs) are routinely invoked to make sense of ambiguous, pluralistic, and value-laden terms, across settings ranging from casual interactions to high-stakes institutional applications. A parent might ask a model whether a movie is ``appropriate'' for their child, or a busy employer might request help identifying the most ``qualified'' candidate from a stack of resumes. In each case, the model is invited to settle  terms that do not admit a single, unique interpretation. While such uses may be informal and under-documented, more structured deployments are rapidly emerging. In 2023, an Iowa school district --- seeking to comply with a law prohibiting library books that depict ``sex acts'' --- turned to ChatGPT to screen its entire catalogue rather than review each book manually~\citep{iowasexacts}. Platforms now routinely rely on LLMs for content moderation, tasking them with interpreting terms like “hate speech” or “incitement” \cite{openai2023moderation, techpolicy2025moderation}. AI labs train models to self-regulate based on constitutional-style documents full of deliberately ambiguous principles \cite{bai2022constitutional, ganguli2023capacity}. Governments are piloting LLM tools to parse regulations and provide tailored legal guidance \cite{Blank_Osofsky_2025}. Even judges have begun using these models to assist with legal interpretation \cite{Snell2024, ross2025dc, deleon2024eleventh}. Across these settings, LLMs  have become a new kind of interpretive infrastructure, effectively positioning them as arbiters of meaning.

This paper introduces and theorizes \textit{ambiguity collapse} as a distinct phenomenon that emerges when LLMs are used in such interpretive capacities. We use the term to describe cases in which a model encounters an ambiguous term --- one that genuinely admits multiple, legitimate interpretations --- but nevertheless selects a single meaning, bypassing the human processes of deliberation, negotiation, and contestation which traditionally underpin these interpretive tasks. Ambiguity collapse is thus a form of interpretive closure, in which a model converts open-textured, plural, and indeterminate concepts into precise, explicit, and singular resolutions. Across law, sociology, communication, STS, education, art, and linguistics, ambiguity is often treated as a productive resource: it enables adaptability, invites reflection, sustains contestability, and holds space for competing values. Different disciplines have built practices and institutions around managing, staging, or preserving ambiguity. When LLMs collapse ambiguity rather than support the interpretive practices that make it meaningful, they may produce distinctive epistemic harms, which we illuminate through a taxonomy grounded in the diverse functions that ambiguity serves across disciplines. The taxonomy focuses on what is lost when LLMs resolve ambiguity for us at three levels: process (foreclosing opportunities to deliberate, develop cognitive skills, and shape contested terms), output (distorting the
concepts and reasons agents act upon), and ecosystem (reshaping shared vocabularies, interpretive norms, and how concepts evolve
over time).

Ambiguity collapse is not inevitable; it is driven by identifiable pressures embedded in how these systems are trained, aligned, deployed, and interacted with. Nor do we suggest that every instance of disambiguation is undesirable; models frequently must select a single interpretation to be useful. Ambiguity collapse becomes concerning only if it leads to one of the harms detailed in the taxonomy. Our contribution is conceptual and diagnostic: we identify a pattern that is emerging across domains, explain why it is problematic, and trace its contours through illustrative examples. This, in turn, informs potential mitigation strategies. The paper proceeds as follows. The next section defines ambiguity, and situates our contribution within existing literature. Section 3 introduces our taxonomy of epistemic risks. In Section 4, we turn to three case studies: LLM self-alignment through constitutional-style rules, and LLMs as quasi-adjudicative decision-makers (“LLM-as-a-judge”), and LLMs used for legal interpretation, which illustrate that the risks in our taxonomy are already materializing in high-stakes domains. We conclude  in Section 5 by unpacking the design and deployment pressures that drive ambiguity collapse and sketching mitigation strategies that target those pressures.

\section{Definition, Scope \& Prior Work}\label{sec:Definition, Scope & Prior Work}

\subsection{Definition of Ambiguity} 
Across disciplines, ambiguity has been defined in many different, sometimes incompatible, ways~\citep{senet2021ambiguity}. We use it as an umbrella term for two familiar forms of indeterminacy often distinguished as \textit{ambiguity} and \textit{vagueness}~\citep{Poscher2012AmbiguityVagueness}. \textit{Ambiguity} involves terms that have multiple, discrete meanings: ``harmful'', for example, could mean physical, emotional, or reputational harm, among other meanings~\citep{senet2021ambiguity}. \textit{Vagueness}, by contrast, concerns a single meaning with fuzzy or indeterminate boundaries: ``tall'', for example, clearly picks out a single property, but there is no specific height at which someone becomes tall~\citep{Sorensen2022VaguenessSEP, Endicott2000VaguenessInLaw}. For ambiguous terms, the interpretive challenge is selecting among discrete meanings; for vague terms, it lies not in ascertaining the meaning of a term in the abstract, but in its application --- in whether a particular case falls within its definition~\citep{Solum2010InterpretationConstruction}. Some terms are both ambiguous and vague. Consider how we noted harmful is ambiguous because it admits several distinct conceptions. Even once we fix one conception, say, emotional harm, there remains vagueness in its application, as it is often indeterminate whether particular cases satisfy the principle. Because these two forms of indeterminacy are deeply intertwined, and thus most of the epistemic risks we describe in our taxonomy apply to both, much of our discussion does not sharply differentiate between ambiguity and vagueness. Finally, we cast out of scope of our inquiry cases where a term has multiple interpretations but there is clearly one correct sense of a term. Thus, we do not address the decades-old problem of word-sense disambiguation in NLP, which targets identifying the intended meaning when only one is contextually appropriate~\citep{Navigli}.

\subsection{Distinctiveness of Ambiguity Collapse by LLMs} 
Humans and institutions collapse ambiguity all the time; why, then, treat LLM-driven ambiguity collapse as a distinctive phenomenon? We think LLM-driven ambiguity collapse differs in three respects. First, LLMs are always available, cheap, and frictionless to query, so users route a wide range of questions through the same system. Thus, any tendency to disambiguate is applied at scale~\citep{HowPeopleUseChatGPT, Atlantic2025OutsourcingThinkingAI}. Second, human answers are perhaps more likely to be treated as inputs into one's own reasoning, as prompts for further reflection, conversation, or disagreement. By contrast, model outputs are often treated as \textit{the} answer: a polished ``final'' output that arrives in confident prose~\citep{Anderl2024Conversational, Goddard2012Automation}. Third, institutional disambiguation typically occurs through recognizable procedures, with traceable reasoning and  identifiable authors (e.g., interpretive canons in law, medical clinical guidelines, etc.); their decisions can be contested, appealed, or countered by rival authorities. LLM-mediated disambiguation typically lacks those procedural and institutional safeguards. 

\subsection{Prior Work} 
Our analysis connects to several bodies of prior work. First, this paper contributes to a growing effort to document the epistemic risks posed by LLMs. Recent taxonomies of epistemic harms have mapped the many ways in which LLMs can distort inquiry and mislead users \citep{Coeckelbergh2025AIEpistemicAgency, CoeckelberghDemocracy, Mollema2025TaxonomyEpistemicInjustice, MesseriCrockett2024IllusionsUnderstanding, fazelpour_magnani_2025_aspirational_affordances, DeProostPozzi2023ConversationalAI, KayKasirzadehMohamed2025EpistemicInjusticeGenAI, BurtonEtAl2024CollectiveIntelligenceLLMs, Simon2025QuadrupleDeceptionTrust, HanniganMcCarthySpicer2024Botshit, CriticalThinking, Jakesch}. We situate ambiguity collapse in this broader landscape: a distinctive risk that raises concern about how AI systems shape the processes of and the conditions for knowledge. Second, we draw on interdisciplinary literature that examines the constructive role ambiguity plays in human reasoning and social life. Across a variety of disciplines, ambiguity is treated as a generative feature~\citep{Endicott2000VaguenessInLaw, Hart2012Concept, Waldron2002RuleOfLawFlorida, Eisenberg1984AmbiguityStrategy, Haraway1988SituatedKnowledges, Harding1995StrongObjectivity, Laclau2005OnPopulistReason, gallie1955essentially, graeber2012dead, Empson1930Seven, Iser1972ReadingProcess, eco1989open}. We rely on these traditions' insights to clarify why ambiguity collapse is troubling. Third, our work extends earlier analyses of how technological infrastructures handle ambiguity. Long before LLMs, scholars documented how expert systems, statistical models, and machine-learning classifiers struggle with phenomena that resist crisp categorization~\citep{Hildebrandt2020CodeDrivenLaw, bowker2000sorting, StarGriesemer1989BoundaryObjects, scott2020seeing, fourcade2024ordinal, merry2016seductions, Kristofik2024IndeterminacyLegalLanguage}. We view LLMs as part of this lineage, but we focus on the distinctive ways ambiguity collapse manifests in generative systems. Unlike earlier technologies, LLMs can effectively fix the meaning of terms autonomously. This creates new forms of distortion and new sites of epistemic risk. Finally, our project relates to work on pluralistic AI and efforts to incorporate diverse values, perspectives, or normative commitments into model design~\citep{fazelpour2022diversity,pluralisticalignment, feng-etal-2024-modular,kasirzadeh2024plurality,fazelpour2025value}. Although that literature typically centers on political, cultural, or moral pluralism, our paper focuses on conceptual pluralism at the level of meaning. 

\section{Taxonomy}\label{sec:taxonomy}

Now, we outline our taxonomy of epistemic risks associated with ambiguity collapse. For each harm, we: (i) describe the harm in general terms, (ii) use non-LLM analogues to clarify its logic and situate it within broader epistemic or sociotechnical patterns, and (iii) show how this materializes in the context of LLMs. Table 1 provides a consolidated overview of all risks and illustrative examples, including several additional examples not discussed in the main text. Not every harm will arise in every case of ambiguity collapse, and the categories are not mutually exclusive; some overlap or compound one another. While we focus on LLMs, most risks generalize to other generative AI systems.

\begin{table*}[!tbp]
\centering
\normalsize
\setlength{\tabcolsep}{4pt}
\renewcommand{\arraystretch}{1.05}

\begin{adjustbox}{width=1.00\textwidth,center}
\begin{tabular}{@{}l l p{0.32\textwidth} p{0.48\textwidth}@{}}
\toprule
Level & Harm & Description & Real-World Example \\
\midrule

Process & Deliberative Closure &
The loss of individual deliberative reasoning (e.g., inquiring, reflection and exploration of alternative interpretation) &
Museum chatbots which risk giving visitors definitive interpretations of artworks instead of prompting individual reflection \citep{wangmuseumchatbots, KuciaMuseumChatbots}; LLM-assisted qualitative research workflows that may shift interpretation work like thematic coding away from iterative deliberation over the schema \citep{braun2006using, khan2024automating, wen2025leveraging, dai2023llm, zhang2023redefining}. \\
\addlinespace[2pt]\cmidrule(l){2-4}

& Pedagogical Erosion &
The erosion of cognitive and intellectual skills and habits &
LLM-assisted learners exert less cognitive effort, weaker reasoning and thinner justifications \citep{stadler2024cognitive, kosmyna2025your, fan2025beware, lehmann2024ai, wecks2024generative, bastani2024generative}; a Khan Academy chatbot treats the green light in \emph{The Great Gatsby} as having a single, settled meaning \citep{KhanAICouldSaveEducation}. \\
\addlinespace[2pt]\cmidrule(l){2-4}

 & Displacement of Interpretive Authority &
A shift in the power to settle meaning from publicly accountable interpretive actors to upstream model designers and deployers &
Judges relying on LLMs to determine the ``ordinary meaning'' of legal terms \citep{Snell2024, ross2025dc, deleon2024eleventh}; proposals for ``silicon reasonable persons'' and LLM juries to stand in for humans \citep{arbel2025silicon, UNCLaw2025AIJuryMockTrial}; Grok 4 outputting answers that track Elon Musk’s posts on X \citep{APNewsGrok4xAIColossus}. \\

\midrule

Output & Epistemic Narrowing & 
The contraction of the user’s accessible information space via (i) the model failing to surface multiple reasonable interpretations (loss of alternatives) (ii) the model forcing genuinely borderline or misfitting cases into a crisp label, making the gray zone disappear (loss of residuals) &
\textit{Loss of Alternatives}: When prompted about “democracy” or “free speech,” LLM outputs converge on a small set of definitions, crowding out other reasonable interpretations from the user’s output space \citep{wright2025epistemic};

\textit{Loss of Residuals}: An Iowa school district using ChatGPT to classify books  into ``sex acts / not sex acts'' labels \citep{iowasexacts}; toxicity classifiers that must tag speech as either ``offensive'' or ``not'' \cite{lu-etal-2025-llm}; the IRS Interactive Tax Assistant forcing complex fact patterns into yes/no eligibility boxes \citep{YJREGAutomatedLegalGuidanceResponse}; diagnostic algorithms that map atypical patients into ill-fitting disease categories \citep{DeProostPozzi2023ConversationalAI}; name-based gender prediction systems with only ``male'' or ``female'' outputs \citep{you-etal-2024-beyond}.  \\
\addlinespace[2pt]\cmidrule(l){2-4}

& Normative Smuggling & 
The replacement of the user’s reasons with the model’s normative criteria in disambiguation, coupled with the presentation of the result as neutral or inevitable, masking the underlying value judgment &  
Different LLMs embedding incompatible definitions of ``hate speech'' into moderation decisions, substituting their own conceptual choices for users’ \citep{melis-etal-2025-modular, fasching-lelkes-2025-model}; LLMs enact their own value trade-offs rather than mirroring human preferences \citep{arunasalam-etal-2025-implicit}; LLM-based hiring tools that invite employers to describe in broad terms the candidates they want, and then return shortlists and scores \citep{JuiceboxAI, FountainAIRecruiter}. \\

\midrule
Ecosystem & Interpretive Lock-In &
The entrenchment of a model’s initial interpretation as the default in downstream use when its outputs are reused, gradually crowding out alternative meanings &
LLMs used as annotators whose labels become entrenched as ``ground truth'' in benchmark datasets \citep{schroeder-etal-2025-just}; Text-to-image pipelines expand prompts into a fixed “system representation”; circulation of resulting images entrenches that early framing  \citep{fazelpour_magnani_2025_aspirational_affordances}. \\
\addlinespace[2pt]\cmidrule(l){2-4}

& Monoculture &
Uniformity in how ambiguous expressions are resolved across models, reducing interpretive diversity at the ecosystem level &
When asked to name “prominent figures” across fields, LLMs converge on the same small set of names across languages \citep{goethals-rhue-2025-one}; LLMs disproportionately reflect the opinions of particular demographic and ideological groups \citep{santurkar2023whose}. \\
\addlinespace[2pt]\cmidrule(l){2-4}

 &  Breakdown of Shared Meaning &
The degradation of shared meaning when model-mediated disambiguation either (i) yields systematically different resolutions of the same term across groups (miscommunication) or (ii) fixes a term whose openness was enabling coordination (coalition fracture) &
\textit{Miscommunication}: Ideologically tuned LLMs that give systematically different answers about the same political topics to users in different echo chambers \citep{Sharma};

\textit{Coalition Fracture}: ML models that force policymakers to pick a single, explicit objective when multiple goals previously coexisted \citep{coyle2020explaining}; LLM-based ``Fatwa Engines'' \citep{AIMuFTI} or ``Digital Rabbis'' \citep{DigitalRabbi} that collapse plural rich religious traditions into a single meaning \citep{atif2025sacred}; Retrosynthesis ML models resolve ambiguous criteria (e.g., “similarity”) by fixing a single operational definition, potentially displacing plural, context-sensitive lab norms that previously enabled coordination among chemists \cite{BlackshawBlackshaw2025Algorithms}. \\
\addlinespace[2pt]\cmidrule(l){2-4}

 & Diminished Tolerance for Ambiguity &
The collective weakening of capacity to sit with ambiguity via habitual delegation of meaning-making to models &
Popular accounts warning that everyday reliance on AI assistants may deskill users and make them less comfortable with uncertainty and open-ended problems \citep{Atlantic2025AIDeskilling, Atlantic2025OutsourcingThinkingAI}. \\

\bottomrule
\end{tabular}
\end{adjustbox}

\caption{Risks of ambiguity collapse at the process, output, and ecosystem levels, with illustrative examples.}
\label{tab:ambiguity-collapse-risks}
\end{table*}

\subsection{Process-Level Risks} 
Process-level risks capture what may be lost when LLMs resolve ambiguity in our stead – foreclosing opportunities to deliberate, to develop cognitive skills, and to participate in shaping how contested terms are understood.

\subsubsection{Deliberative Closure:}
the loss of individual deliberative
reasoning (e.g., inquiring, reflection
and exploration of alternative interpretation).

Ambiguity often induces deliberation: it forces people to inquire, compare competing interpretations, and confront the tradeoffs that different resolutions would impose~\citep{Shiffrin2010OccasionalVirtuesFog}. In doing so, it can support reflection and value formation. 
Nguyen's account of \textit{value capture} shows how relying on crisp institutionalized metrics leads us to outsourcing the work of deliberating about our values~\citep{Nguyen2024-NGUVCH}. He describes how before U.S. law school rankings became dominant, applicants typically spent significant time reading about schools' missions, visiting campuses, and comparing distinct institutional identities, forming an internal sense of what they valued in a legal education. Once a single, simplified metric became the default, applicants could lean on the ranking instead of asking themselves what “best” actually meant to them. Similarly, in legal and political domains, ambiguity can keep people engaged as active participants in meaning-making rather than passive recipients of rules. Shiffrin makes this point in an intentionally counterintuitive context: speed limits. Even in domains like traffic, where we often think we want crisp rules, some legal “fog” can improve both safety and responsibility, because it keeps people attentive to context rather than complacently relying on rote compliance with fixed signals ~\citep{Shiffrin2010OccasionalVirtuesFog}. She points to evidence from “shared space” design suggesting that making uncertainty salient (e.g., reducing overly specific traffic markings/signals) can reduce speeds and accidents by prompting drivers to pay closer attention to nearby cars and pedestrians and actively negotiate what safety requires in the moment.

LLMs can undermine these virtues when introduced into analogous settings: for instance, museum pilots that use chatbots to answer visitor questions about artworks, rather than inviting museum guests to grapple with the meaning of art themselves \citep{wangmuseumchatbots, KuciaMuseumChatbots}. A similar loss occurs when LLMs are used in qualitative research. In thematic analysis, coding is meant to be an iterative, interpretive practice: researchers sit with the data, argue over awkward extracts, and even park unruly codes in a temporary “miscellaneous” theme so they can return to them later \cite{braun2006using}. Deviant or hard-to-place cases are analytically valuable, prompting teams to revisit their categories and refine what their themes capture. When LLMs are used to auto-code or pre-tag texts \cite{khan2024automating, wen2025leveraging, dai2023llm, zhang2023redefining} that deliberative work can be displaced.

\subsubsection{Pedagogical Erosion:}
the erosion of cognitive and intellectual skills.

Independent of its deliberative virtues, ambiguity also carries pedagogical value. Working through ambiguity activates reasoning, discernment and judgment -- capacities that are critical to learning. 
The learning sciences have long emphasized that difficulty and uncertainty improve understanding: “desirable difficulties” \citep{bjork2020desirable} and “productive failure” \citep{kapur2008productive} describe intentionally effortful or underspecified tasks whose struggle strengthens retention, transfer, and conceptual understanding. Dewey similarly argues that thought begins not in knowledge but in  ambiguity, or “forked-road situation”, where familiar habits fail, and the learner must arrive at a solution through reflective inquiry \citep{dewey2022we}. The ability to remain in that state of suspended judgment is, for Dewey, the foundation of intelligence. Across domains, much of what we call learning consists of this kind of work.

This dynamic maps onto common uses of LLMs. In education, echoing worries about AI-driven “de-skilling,” recent work finds that LLM-assisted learners often exert less cognitive effort and, as a result, display weaker reasoning, thinner justifications, and shallower engagement with underlying concepts \citep{stadler2024cognitive, kosmyna2025your, fan2025beware, lehmann2024ai, wecks2024generative, bastani2024generative}. An example from a Khan Academy tutoring chatbot tutorial on The Great Gatsby gives a sense of what can be lost: when asked about the green light at the end of Daisy’s dock --- a multifaceted and ambiguous metaphor --- the model replies that it “represents my yearning for the past and my hope to reunite with Daisy” \citep{KhanAICouldSaveEducation}. The green light is treated as something settled instead of an open symbol students can linger over and learn to make sense of for themselves. 

\subsubsection{Displacement of Interpretive Authority:}
A shift in the power to settle meaning from publicly accountable actors to upstream model designers and deployers.

To disambiguate evaluative, socially contested, or politically charged concepts can be thought of as an exercise of power \citep{Laclau2005OnPopulistReason, {haslanger2020going}}. For \textit{essentially contested concepts} \citep{gallie1955essentially} (“democracy”, “justice”, “fairness”) whose reference is temporarily undecided across rival interpretive communities, fixing meaning is an attempt to claim definitional authority, privileging one reading while sidelining others \citep{Gadamer2004TruthMethod}. In judicial, legislative, journalistic, and educational contexts, such authority is exercised by identifiable actors who must justify their interpretations and can be held accountable, or left open to public contestation. When LLMs fix meaning, this authority is outsourced to LLMs. 
Epistemically, this shift can amount to  epistemic exclusion or participatory injustice \citep{Dotson2014EpistemicOppression, Hookway2010-HOOSVO}: barring people from contributing to meaning-making and thereby excluding them from knowledge production. Furthermore, when LLMs are the default arbiters of meaning, human voices may be accorded diminished credibility \citep{Fricker2007EpistemicInjustice}, potentially leading to testimonial smothering: when those repeatedly disadvantaged in communicative exchanges preemptively withhold their perspectives, anticipating that their contributions will be dismissed.

We already see LLMs beginning to wield authority in consequential settings. Judges have deployed LLMs to probe the “ordinary meaning” of terms, granting systems with no democratic pedigree a role in interpreting law \citep{Snell2024, ross2025dc, deleon2024eleventh}. Elsewhere, experiments with “LLM jurors” and “silicon reasonable persons” explore using LLM ensembles to stand in for a jury of one’s peers or for the legally “reasonable” person in applying standards of care or responsibility \citep{arbel2025silicon, UNCLaw2025AIJuryMockTrial}. An even starker illustration comes from xAI’s Grok 4: reporters have found that, when asked about divisive political topics, the model sometimes searches platform X for Elon Musk’s posts and treats his views as guidance for its own answers \citep{APNewsGrok4xAIColossus}. In each of these cases, authority over contested meaning migrates from publics, juries, and affected communities to privately governed systems, and, at the limit, to the idiosyncratic judgments of a single platform owner.

\subsection{Output-Level Risks} 

Output-level harms capture how disambiguation may distort the underlying concept itself leaving agents acting on deficient or mis-framed reasons, thereby undermining their epistemic agency.

\subsubsection{Epistemic Narrowing}

the contraction of the user’s accessible information space via (i) the model failing to surface multiple reasonable interpretations (loss of alternatives) (ii)
the model forcing genuinely borderline or misfitting cases into a crisp label, making the gray zone disappear
(loss of residuals).

(i) \textit{Loss of Alternatives}. When the model selects one interpretation, the user loses insight into all reasonable alternative interpretations of a particular concept. For instance, if a user asks about the meaning of “privacy”, they may benefit from learning about distinct frameworks (control vs. contextual integrity vs. freedom from surveillance). The stakes are especially high for complex and multifaceted concepts. For example, \textit{wicked problems} \cite{RittelWebber1973DilemmasPlanning}, such as climate change, admit no definitive formulation; \textit{essentially contested concepts} \cite{gallie1955essentially}, such as justice or democracy, derive their meaning through ongoing collective argument. These domains are entangled with conflicting values in ways that make sustained pluralism over problem framings epistemically valuable. That pluralism, in turn, organizes inquiry -- shaping what counts as  harm, which tradeoffs matter, what evidence is probative, and who the relevant stakeholders are. Collapsing that plurality limits what the user sees and draws on in their reasoning, impoverishing inquiry. Wright, for example, shows that models offer only a small set of generic descriptions for value-laden concepts such as “democracy” and “free speech”, while rarer but still reasonable interpretations of the terms effectively disappear from the output space \citep{wright2025epistemic}.

(ii) \textit{Loss of Residuals}. A second route that might lead to contraction of the user’s information space is loss of residuals: the model forces genuinely indeterminate cases into rigid categories, making the gray zone itself disappear. We can distinguish two kinds of residuals: (a) cases that sit between categories, where multiple classifications are defensible; and (b) cases that fall outside the schema altogether, where no available label fits. Residual categories mark the limits of a classification system’s descriptive reach, and with it, the erasure of the social realities that resist its conceptual frame \citep{StarBowker2007EnactingSilence}. Losing sight of residuals makes the underlying gray zone illegible: users no longer see the misfitting details that spill out of the available categories, and the strain those details place on the existing classificatory frame. LLMs are routinely pushed toward decisiveness even when a case is gray: forcing gender-neutral names to “male/female” \citep{you-etal-2024-beyond}, slotting atypical patients into ill-fitting diagnostic boxes \citep{DeProostPozzi2023ConversationalAI}, or confidently tagging text as offensive or not even where human annotators sharply disagree \citep{lu-etal-2025-llm}. Sometimes this is a function of  deployment context, or imposed by the users themselves \citep{rottger-etal-2024-political, 10.1145/3715275.3732147, li-etal-2025-decoding-llm}. But even when no explicit schema is provided, LLMs are trained to appear helpful, confident, and decisive, which may bias them toward picking a salient, entrenched category rather than expressing uncertainty~\citep{ZhouEtAl2024LessReliable}. Returning to the Iowa school that used ChatGPT to identify books containing “sex acts” \citep{iowasexacts}: A text like the Bible most certainly contains passages that gesture toward sex acts; it is a classic borderline case. When the model returns a “yes” or “no”, the book is absorbed into a clean category, and its borderline status, as well as its features that fall outside the scheme, drop out of view. A different example comes from Blank and Osofsky’s study of the IRS’s automated Interactive Tax Assistant, which shows how binary question trees (“yes/no” on whether an expense is “necessary” etc.) force complex, atypical fact patterns into pre-coded templates \citep{YJREGAutomatedLegalGuidanceResponse}. Taxpayers may rely on that answer only to learn later that courts disagree or that they were never entitled to the deduction. In both settings, what a user might need is a brief map of the gray zone – why the case cuts both ways – rather than a binary “yes” or “no”.

\subsubsection{Normative Smuggling:} The replacement of the user’s reasons with the model’s normative criteria in disambiguation, coupled with the presentation of the result as neutral or inevitable, masking the underlying value judgment.

When a model chooses among multiple interpretations, it may select one that the user would not have chosen themselves. Just like other simplified institutional metrics \citep{Nguyen2024-NGUVCH}, model outputs often embody external criteria rather than the user’s situated interpretive standpoint. Had users reflected for themselves, they might have reached conclusions better suited to their goals; instead, they inherit interpretations built for an “average” user and lose the capacity to tailor meaning to their own lives which is central to “well-being and flourishing". Arunasalam et al. show that across a battery of mundane, subjective choice tasks, LLMs systematically enact their own value trade-offs rather than mirroring human preferences \citep{arunasalam-etal-2025-implicit}; and Melis et al. and Fasching et al. find that different models encode distinct implicit definitions of “hate speech”, which may diverge from the contextual aims or values of those who rely on them \citep{melis-etal-2025-modular, fasching-lelkes-2025-model}.

A further layer of harm arises when this value-ladenness is concealed. Bell’s account of descriptive masquerade captures this dynamic ~\citep{Bell2024-BELQTF}. Descriptive masquerade refers to thick concepts that present themselves as if they were purely descriptive, while concealing that a value judgment is doing the work. Obesity, for example, is routinely treated as a neutral descriptor even though it carries entrenched moral and social evaluations about discipline, responsibility, and worth. In Bell’s terms, the concept “pretends only to describe” while quietly wrapping evaluative judgments. We extend this analogy to our setting to describe cases where the model selects a single interpretation and presents it as the inevitable, determinate meaning, as if no alternative readings were possible. As Bell notes, this both obscures the fact that a choice existed and bolsters the authority of the choice that was made, shielding the underlying value-judgment from scrutiny. Consider a hiring manager who asks an LLM which candidate is the most qualified for a particular role from a stack of CVs. The term \textit{qualified} is normatively contested: it could mean productivity, team cohesion, leadership skills, etc. \citep{IndeedGoodEmployeeQualities}. The model will not necessarily surface this range of possibilities. Instead, it might operationalize \textit{qualified} in terms of whatever proxies it treats as salient, and return a confident answer naming a particular candidate, as though there were a ground truth about who the qualified candidate is. A manager who relies on the output may be  under the impression that they are merely deferring to an objective assessment. Emerging LLM-powered “AI recruiters”, which invite employers to describe, in broad natural-language prompts, the kind of candidate they want – “high-potential leaders”, “strong culture fit” “top talent” – and then return shortlists and scores, make this risk concrete \citep{JuiceboxAI, FountainAIRecruiter}.

\subsection{Ecosystem-Level Risks} 

Ecosystem-level risks arise when ambiguity collapse scales: individual disambiguation decisions accumulate to reshape shared vocabularies, interpretive norms, and how concepts evolve over time.

\subsubsection{Interpretive Lock-In:} the entrenchment of an early resolution of an ambiguous concept as the
default meaning in downstream use.

Resolving ambiguity involves choosing one among several meanings, setting aside alternatives that previously fit. Once that choice circulates, through documents, tools, or discourse, it can harden into a convention. Reopening the question later becomes costly: people have internalized the frame, and institutional routines and practices now depend on it.
Lock-in is a problem that plagues other domains, though some have a particular way to manage it. Legal adjudication offers a useful analogy. Law must preserve contestability, yet it is path-dependent: once a court settles on an interpretation, precedent constrains later courts. Precisely because of this, judges and legal scholars developed techniques that allow for conceptual development before ossifying into rigidity. Some judges follow decisional minimalism: deciding cases narrowly (addressing only the specific dispute) and shallowly (avoiding theoretical pronouncements) so that foundational questions remain open for future contestation and deliberation \citep{sunstein110supreme}. Moreover, courts can “decide not to decide” when reaching the merits would freeze a live controversy prematurely \citep{Bickel1961}. Doctrines like standing, ripeness, and mootness allow the court to postpone decision until broader political or doctrinal conversations have matured.

LLMs, too, can precipitate premature interpretive lock-in. In literary translation, if a model renders a deliberately ambiguous and layered phrase into one fixed expression, the nuance of the original may disappear; if that version is adopted in subsequent editions, the interpretation may become canon, leaving little room to reintroduce the ambiguity later \citep{Heuser2025GenerativeAesthetics}. Or, when LLMs are used as annotators in social science, labeling data along fuzzy dimensions – say, whether a post is “toxic” or “pro-democratic” – their particular resolution of those ambiguities becomes baked into the dataset, shifting the distribution of labels and shaping downstream analyses that treat those categories as ground truth \citep{schroeder-etal-2025-just}. 

\subsubsection{Monoculture:} uniformity in how ambiguous expressions are resolved across models, reducing interpretive diversity at the
ecosystem level.

Sometimes, all systems may collapse ambiguity identically. This uniformity may emerge from the concentration of resources required to develop such models, resulting in a landscape dominated by a handful of well-resourced actors \citep{korinek2025concentrating}. At the same time, these models are designed to serve as infrastructure on which a range of downstream applications are built. As more systems are built from the same models their outputs may converge \citep{10.5555/3600270.3600535}. The epistemic cost of such convergence is the suppression of interpretive plurality. When many systems collapse contested concepts in the same way, alternative conceptions are gradually crowded out. Interpretive monoculture also interacts with lock-in: conceptual progress requires a plurality of frames against which ideas can be compared, challenged, and revised. As divergent conceptions disappear, the epistemic ecosystem loses the friction and raw material from which new distinctions and more refined understandings can develop. Over time, this homogenization may produce an illusion that ambiguity never existed – that there is only one correct or self-evident meaning: users may come to inhabit an informational environment that feels more coherent and settled than it really is. 

When models collapse ambiguity not only uniformly but in a way that privileges outlooks of certain demographic or ideological groups, this can amount to \textit{hermeneutic capture}. President Trump’s recent executive orders demanding AI systems “free from ideological bias” and attacking so-called “woke AI” can be read as expressions of this anxiety. A growing body of research shows that the alignment process can embed specific political or ideological orientations, diverging from the distribution of views in the broader population \citep{santurkar2023whose, alkhamissi-etal-2024-investigating, TaoEtAl2024CulturalBiasLLMs, QuWang2024PublicOpinionSimulation}. And at most extreme end of this trend, as noted above, reports on xAI’s Grok 4 suggest that its answers to divisive geopolitical questions sometimes anchor themselves to Elon Musk’s posts on X, effectively tethering its interpretations to a single powerful actor \citep{APNewsGrok4xAIColossus}. Taken together, this skews the shared interpretive resources that people rely on to make sense of the world, privileging one set of values and ideologies while sidelining others.


\subsubsection{Breakdown of Shared Meaning:} the degradation of shared meaning across groups, either through the (i)
fragmentation of a single term into divergent meanings (miscommunication); or by (ii) fixing a term whose openness was doing coordinating work (coalition fracture).

(i) \textit{Miscommunication}. when a model resolves the same ambiguous expression differently for different users, disambiguation  can become a source of misunderstanding. Users may continue invoking the same label – believing they are aligned – while each is operating with a distinct interpretation \citep{mustajoki2012speaker}. LLMs often reach divergent outcomes when interpreting the same term \citep{uceda2024reasoning, purushothama2025not, WaldonEtAl2025LLMLegalInterpretation, choi2025off}, and when disambiguation is filtered through ideological echo chambers, the selected meaning reinforces the user’s epistemic environment \citep{Sharma}. As these divergent interpretations circulate, they can generate systematic unclarity and encourage unjustified inferences. Imagine a model that disambiguates the term “equality” differently for different users, each time selecting one conception – formal equality, substantive equality, equality of opportunity, and so on. Downstream, when invoking equality, users may assume that others share their meaning but instead talk past one another, drawing inferences or offering justifications that rest on incompatible conceptual foundations \citep{SteelEtAl2018MultipleDiversity, mulligan2019thing}. This dynamic is not unique to LLMs. Within the AI research community many core terms -- machine unlearning, bias, mode collapse, and interpretability -- are used in multiple, partly incompatible senses. Several researchers \citep{cooper2024machine, blodgett-etal-2020-language, schaeffer2025position, lipton2018mythos} have called attention to how shifting definitions impede cumulative progress: it is difficult to know what claim is being advanced, what evidence supports it, or whether researchers are addressing the same phenomenon. LLMs can amplify this instability. As users increasingly rely on LLMs to interpret or operationalize these terms, models risk hardening inconsistent or idiosyncratic readings, resulting in unjustified cross-concept inferences, diminished shared understanding, and the obscuring of the normative commitments embedded in each interpretation.

(ii) \textit{Coalition Fracture}. Disambiguation can also destroy coordination that depends on managed interpretive openness. There are settings in which ambiguity sustains coordination across difference \cite{haslanger2020going}. Social movements rely on slogans like “we are the 99\%” precisely because they are thin on detail, allowing heterogeneous grievances to gather under a single chant (e.g., debt, housing, wages, and political capture). Diplomats similarly draft deliberately open clauses to enable agreement among states with divergent interests \citep{Eisenberg1984AmbiguityStrategy}. Star and Griesemer’s notion of \textit{boundary objects} captures the same pattern: shared categories whose looseness lets different communities coordinate around a common term without erasing their substantive differences \citep{StarGriesemer1989BoundaryObjects}. In all of these cases, ambiguity functions as an epistemic affordance for groups: it furnishes a shared but flexible reference point that different parties can use to reason together, organize, and maintain fragile coalitions. When a system disambiguates or concretizes the shared term, it fixes meaning in one direction, revealing that parties were not “talking about the same thing” after all, and thereby collapsing the coalition or cooperation that the ambiguity had made possible. LLMs are also being introduced into domains whose internal stability depends on managed interpretive plurality. Coyle and Weller highlight how the deployment of machine learning in policy contexts like criminal justice, which require ambiguous objectives to be resolved unequivocally, forcing policymakers to be explicit about their objectives, exposing implicit trade-offs and removing the flexibility often required to maintain consensus in complex sociopolitical domains \citep{coyle2020explaining, esposito2022transparency}.

\subsubsection{Diminished Tolerance for Ambiguity:} the collective weakening of our capacity to sit with ambiguity through habitual delegation of meaning-making to models.

Reliance on LLMs may also weaken the interpretive habits required for critical, pluralistic, and reflective reasoning. We have long known that friction-reducing technologies can reshape epistemic capacities. For example, people who rely on search engines retain where to find information rather than the information itself \citep{SparrowEtAl2011GoogleEffects}, and heavy GPS users form weaker spatial memories when navigating on their own \citep{DahmaniBohbot2020GPSMemory}. Nguyen’s account shows a parallel shift in evaluative habits: instead of wrestling with the messy, incommensurable values at stake in choosing a school, neighborhood, or job, people defer to rankings that present those choices as if they lay along a single ordered scale \citep{Nguyen2024-NGUVCH}. It is plausible that an analogous dynamic could unfold with the skill of grappling with ambiguity. When LLMs consistently resolve ambiguity for us, we, as a society, may gradually lose the ability to tolerate the discomfort of uncertainty \citep{Atlantic2025AIDeskilling}. Constant disambiguation encourages over-reliance on premature closure. And like any underused muscle, interpretive capacity weakens through disuse; over time we may become less capable of exercising them at all \citep{Atlantic2025OutsourcingThinkingAI}.

The epistemic stakes are significant. The skills involved in grappling with ambiguity are precisely those that feminist epistemology identifies as constitutive of better knowledge. Harawy’s "situated knowledges" underscores that all knowledge is  partial, and that acknowledging this partiality makes claims more accountable and less prone to domination by a single viewpoint \citep{Haraway1988SituatedKnowledges}. Harding’s "strong objectivity" similarly argues that inquiry becomes more rigorous when it incorporates multiple standpoints, especially those marginalized or excluded, because they can expose hidden assumptions and blind spots \citep{Harding1995StrongObjectivity}. On this view, disagreement and contestation are epistemic assets: they allow differing background beliefs and values to critically interrogate one another. When we lose the capacity to tolerate ambiguity or multiplicity, we risk drifting back toward a thin, pseudo-neutral “view from nowhere”.

\section{Case Studies}\label{sec:Case Studies}
Now, we examine three high-stakes domains to make the practical stakes of ambiguity collapse vivid, and to demonstrate how the taxonomy can be used as a diagnostic lens for identifying the risks at play.

\subsection{Model Self-Alignment}

Aligning LLMs after pre-training ensures they meet criteria like safety and helpfulness. The conventional method, reinforcement learning from human feedback (RLHF), is labor-intensive and does not scale well, as it relies on extensive human annotator rankings \citep{ouyang2022training}. More recent, scalable techniques shift the critique and correction burden to the model itself using natural language principles. These include Constitutional AI, where the model self-polices by critiquing outputs against a set of high-level normative rules \citep{bai2022constitutional}, and moral self-correction at inference time, where the model revises its output based on a prompt \citep{ganguli2023capacity, yuan2025inference, liu-etal-2024-intrinsic}. The appeal of these newer methods is that they allow the model to self-enforce alignment criteria across domains with a fixed set of rules, reducing the need for constant human oversight.

We argue that these alignment approaches are conducive to ambiguity collapse. Take Constitutional AI. To scale across contexts, constitutional rules must be drafted in general, open-textured terms. But breadth comes at the price of indeterminacy: concepts like “harmful”, “aggressive”, or “biased” and even broader directives like “choose the most worthwhile response” or what is “less risky for humanity in the long run”, \citep{huang2024collective} admit multiple reasonable interpretations. Handing such principles to a model  effectively delegates the task of choosing among these competing interpretations to the system itself.
Recent empirical work by He et al. makes this dynamic visible: when they apply a panel of independently trained judge models to Anthropic’s constitutional rules, the models often reach divergent judgments about the same scenarios \citep{he2025statutory}. 
He et al. thus identify “interpretive ambiguity” as a fundamental challenge, but they frame the problem primarily as one of inconsistency. Their proposed solutions – refining rules to make them less open-textured until disagreement disappears, or constraining interpretive strategies so models apply them uniformly – treat divergence as noise.
Our concern is somewhat different. We see divergence as signaling a plurality of reasonable, defensible moral positions.
When 85\% of scenarios produce disagreement on what counts as “humanity’s benefit,” this is a sign that the rule is functioning as a delegated site of normative choice -- one on which reasonable people can and do disagree. Here, that interpretive work is carried out without an identifiable decision-maker or contestable procedure (\textit{displacement of interpretive authority}; \textit{deliberative closure}), and then presented downstream as a determinate verdict that reads like a neutral rule-application, even though it reflects a particular choice among reasonable moral readings (\textit{normative smuggling}). From this perspective, even if every judge model had converged on the same outcome, the underlying concern about ambiguity collapse would remain, albeit concealed. In fact, such convergence might be more troubling, because it would create the illusion of a uniquely correct reading and false consensus (\textit{monoculture}).

We see a similar dynamic at work in moral self-correction. 
Wang et al. show that when prompted to provide “unbiased” answers, models consistently default to difference-unawareness (i.e., erasing existing differences between groups) even though plausible conceptions of fairness require recognizing differences across groups to avoid perpetuating inequality \citep{wang-etal-2025-fairness}. Moral self-correction thus risks collapsing ambiguity by implicitly privileging a particular conception of fairness -- one that may not match the user’s aims or context -- while presenting that choice as if it were the determinate, neutral meaning of “unbiased” (\textit{normative smuggling}). Scaled across deployments, the same implicit resolution can harden into a shared default (\textit{monoculture}).

Despite opposite dynamics --- divergence in one, convergence in the other --- both reflect the same problem: contested normative principles allow multiple legitimate readings, yet models covertly privilege one. Plurality is either exposed as inconsistency or masked by convergence, but in both the resolution lacks visibility, contestability, and accountability.

\subsection{LLM-as-a-Judge}

We use LLM-as-a-judge to refer to settings in which a model is tasked with evaluating outputs against a rubric, acting as scalable, low-cost proxies for humans \citep{GuEtAl2024SurveyLLMasJudge, Zheng, li2024llms}. \footnote{While “LLM-as-a-judge” is often used for LLMs judging model outputs, we use it in a broader sense, to refer to LLMs applying rubrics or policies to assess human or model-generated content.} While this approach first emerged as a way to scale model evaluation, it has expanded far beyond this context: Judge models are now proposed as referees in scientific peer review, graders of student assignments, moderators flagging hate speech or misinformation, assessors of submissions in hiring, as annotators in social-science research, and even as adjudicators in games, among other contexts.

Across these applications, many tasks are inherently subjective: what counts as a “novel” idea, “harmful” speech, or a “strong” job application varies across legitimate viewpoints \citep{pavlick-kwiatkowski-2019-inherent, davani-etal-2022-dealing, comedi-ws-2025-1, min-etal-2020-ambigqa, goyal2022your, ChenZhang2023JudgmentSieve, dsouza-kovatchev-2025-sources}. To operationalize these subjective constructs, in the best case, practitioners supply natural-language rubrics and ask the model to assign a score or label. To illustrate, consider OpenAI's recent proposed a framework for detecting "political bias" in ChatGPT that uses a judge model to grade outputs against a detailed rubric \citep{OpenAI2024PoliticalBiasLLMs}, instructing it to flag issues such as personal political expression, asymmetric coverage, or invalidating the user’s viewpoint, and to assign a bias score. But crucially, many of the key terms in that rubric – “objectivity”, “legitimate viewpoints”, “valid justification”, “appropriate coverage” are themselves multifaceted, highly ambiguous concepts: different political or theoretical traditions understand each of these terms differently. The model must choose one interpretation when applying these criteria.\footnote{We discuss the OpenAI case not to single out any particular actor, but because we think it exemplifies a broader pattern that we expect to arise across many LLM-as-a-judge contexts.}

Some practitioners may attempt to mitigate this by explicitly defining terms that have multiple meanings. And while this addresses \textit{ambiguity}, it does not always eliminate \textit{vagueness}: concepts whose meaning is clear but whose boundaries are fuzzy. Even with a fairly detailed rubric, when a model encounters a borderline case, it must decide whether it fits the criterion \citep{GuerdanEtAl2025ValidatingLLMJudge}. For example, a policy might clearly define “harmful content”, yet still leave a vast indeterminate zone where it is not clear if content is harmful enough to warrant removal. Forcing a determination in genuinely indeterminate scenarios offloads normative judgment about hard cases to the model (\textit{displacement of interpretive authority}); hollows out deliberative work, where grappling with borderline cases is how researchers iteratively refine their codes, questions, and conceptual frames (\textit{deliberative closure}); enforces the model’s latent value commitments, obscuring the normative character of the classification, appearing as though it straightforwardly follows from the rubric (\textit{normative smuggling}); and imposes a categorical label where a more appropriate response might have been surfacing uncertainty or prompted reconsideration of the classification scheme (\textit{loss of residuals}).

\subsection{Legal Interpretation}
In many legal systems, a central task for judges is interpretation: deciding how ordinary readers would understand contested words in statutes and other legal texts. Courts regularly face disputes about what legal texts mean in hard cases -- for example, whether an ambulance counts as a “vehicle” under a park ordinance banning vehicles. Judges, practitioners, and scholars have begun experimenting with LLMs for legal interpretation. On the bench, Judge Kevin Newsom used ChatGPT to test the meaning of terms such as ``landscaping'' and ``physically restrained''. \citep{Snell2024}. In the D.C. Court of Appeals, Judge Joshua Deahl posed a more fact-specific query in an animal-cruelty case – whether leaving a dog in a car for over an hour on a 98-degree day is harmful – using the model to test claims about “common knowledge” \cite{ross2025dc}. Scholars have pushed further, offering systematic frameworks for LLM-assisted interpretation \citep{McAdamsEngel2024, ArbelHoffman2024GenerativeInterpretation, UnikowskyInAIWeTrust, KieffaberGandallMcLaren2025WeBuiltJudgeAI}. Proponents contend that such systems can be more accurate than judges drawing on dictionaries or intuition, by approximating a “majoritarian” reading of text; more consistent, by constraining judicial discretion and dictionary shopping; and more efficient, by making interpretive analysis cheap and scalable [as described by \citep{grimmelmann2025generative}]. At the same time, a growing body of work pushes back against this optimism \citep{grimmelmann2025generative}, citing issues like reliability \citep{choi2025off, WaldonEtAl2025LLMLegalInterpretation}. While we sharethese concerns, we think our lens of ambiguity collapse puts pressure on a slightly different aspect.

Consider Snell, an insurance-coverage dispute that turned on whether installing an in-ground trampoline counted as “landscaping”. Judge Newsom  turned to an LLM to test the “ordinary meaning” of the term. Doing so implicitly assumes that (1) that there is one correct answer; and (2) that a  capable LLM can reliably supply that answer. Much existing criticism targets the second assumption. But ambiguity collapse highlights a problem with the first assumption, even if the second could somehow be fixed.
The term “landscaping” is ambiguous (e.g., aesthetic alteration of landforms, functional maintenance of natural elements, environmental modification, etc). A judge would ordinarily canvass competing definitions, weigh policy considerations, and justify selecting among several plausible interpretations – ideally with a public justification that can be critiqued, appealed, or distinguished in later cases. When asked whether installing a trampoline counts as landscaping, the model responded with an unqualified conclusion – “yes, installing an in-ground trampoline can be considered part of landscaping” (or, in other prompts, the opposite), accompanied by a tidy explanation. The response installs a particular evaluative conception of what the term should capture that may not align with the judges', while formatted as if it were an empirical fact (\textit{normative smuggling}); sidelines the rival readings the dispute turns on (\textit{loss of alternatives}); all while displacing the justificatory practices that anchor legitimacy in law (\textit{deliberative closure}; \textit{loss of interpretive authority}. \citep{pruss2025against, coan2025artificial}.\footnote{We are bracketing  jurisprudential debates about when, if ever, legal meaning is fully determinate. Our aim here is more modest. First, to make explicit the assumptions that are smuggled in when interpretive questions are handed to LLMs: that there is a single correct answer of the right kind, and that it is the model is capable of supplying it. Second, to note that these assumptions should be unsettling even for avowed textualists, who mostly recognize that there are hard cases in which the available legal materials underdetermine the result, leaving judges with responsibility for exercising judgment under conditions of uncertainty.}

\section{Mitigations}\label{sec:Mitgations}

Ambiguity collapse is the product of design choices at multiple layers throughout the pipeline, as seen in figure 1. In what follows, we sketch design principles that could address ambiguity collapse.
\begin{figure}[h]
    \centering
    \includegraphics[width=1.0\linewidth]{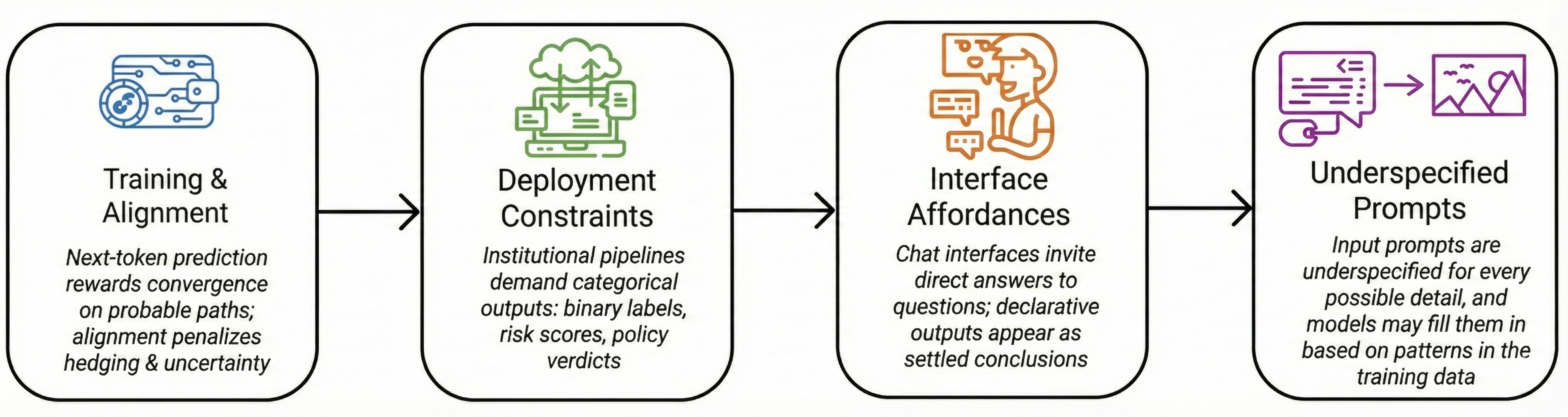}
    \caption{Drivers of ambiguity collapse across the pipeline.}
    \Description{A schematic showing drivers of ambiguity collapse at multiple layers of the pipeline (data/training, model objectives, interface, and deployment), with arrows indicating how pressures at each layer push toward premature meaning-settlement.}
    \label{fig:mitigations}
\end{figure}

\textit{Training \& Alignment.} Training and post-training objectives push LLMs toward a single confident interpretation, penalizing hedging and uncertainty \cite{leng2024taming, Sharma}. Thus, counteracting ambiguity collapse means treating ambiguity management as an explicit objective rather than a defect. One route is to train models to recognize when a query admits multiple readings and respond by surfacing that plurality or asking clarification questions \citep{kim-etal-2024-aligning, lee2023asking}. Ambiguity-focused datasets and benchmarks (AmbigQA \citep{min-etal-2020-ambigqa}, CLAMBER \citep{zhang-etal-2024-clamber}) can be incorporated into instruction-tuning and preference models so that “good” behavior includes enumerating options. Alignment schemes can also adjust rewards so that abstention or hedging is sometimes preferred over decisiveness. Bastani et al.’s “GPT Tutor” show how simple guardrails can steer the same base model toward giving stepwise hints rather than full solutions \citep{BastaniEtAl2025AIGuardrailsHarmLearning}.

\textit{Deployment Constraints}. At present, LLMs are often enlisted as blunt classifiers, much like earlier machine-learning-based predictive models (content-moderation pipelines that demand categorical labels~\citep{openai2023moderation, fasching-lelkes-2025-model}; LLM-as-a-judge settings that force a single winner~\citep{GuEtAl2024SurveyLLMasJudge, Zheng, li2024llms}; or hiring tools that filter for ``qualified'' candidates~\citep{gaebler2024auditing, anzenberg2025evaluating, gan2024application}). However, LLMs can, at least in theory, convey nuance, plurality, and qualification through language. Precisely because of this, they should expand the space of possible workflows that leverage these advantages. Some emerging work points to different possibilities. Wilner describes a content policy pipeline in which the model evaluates cases against multiple paraphrased variations of the same policy. When these variations lead to inconsistent verdicts, the system flags the instance as a borderline case and routes it back to human policymakers: the model’s inconsistency becomes a diagnostic signal that the policy text itself needs clarification or revision \citep{ModeratingAIYouTube2025}. More speculatively, we can imagine pipelines that shift from “rubric application” (top-down) to “analogical reasoning” (bottom-up). Instead of asking a model to apply a fixed abstract definition of “hate speech” to a new post, a system could ask the model to reason across a curated set of precedent-like examples: “Is this new case more like Example A (allowed) or Example B (prohibited), and why?” This approach builds meaning iteratively and contextually through comparison, much like a common-law judge.

\textit{Interface \& Interaction Affordances}. If current chat interfaces afford one-shot question-answering and make determinacy the default epistemic posture \cite{heersmink2024phenomenology, Heersmink2013TaxonomyCognitiveArtifacts, Magnus2025TrustingChatbots, ibrahim2024characterizing}, mitigation requires reconfiguring those affordances so that ambiguity and pluralism become available ways of using the system \citep{jasanoff2007technologies, nielsen2025law}. Interfaces can be designed to surface the multiplicity of plausible continuations the model can generate: by default, outputs might present several candidate interpretations or labels side-by-side, with brief rationales and indications of where they diverge. Some current models already gesture in this direction: ChatGPT and Claude increasingly answer rewriting requests with  alternative phrasings. Interaction flows can invite users to refine or contest answers – e.g., with lightweight controls that toggle between “give me your best guess” and “show me alternatives”, or that let users request a more conservative, permissive, or multi-principled reading. Rather than “spoonfeeding” users with passive follow-ups (e.g., “would you like me to tell you about X?”), which reinforce a dependency on the system’s initiative, system messages and UI cues can frame responses explicitly as proposals or drafts for human consideration, and can encourage follow-up questions (“What else might this mean?”, “How would this change under a different standard?”) \citep{SarkarHowToStopAICriticalThinking}. These can push users toward exploratory and deliberative uses of LLMs rather than treating them as frictionless substitutes for judgment. 

\textit{Underspecified Prompts}. A subtler driver of ambiguity collapse stems from underspecification \cite{JMLR:v23:20-1335}. Input prompts are necessarily underspecified for every possible detail, and models may fill them in based on patterns in the training data. In image generation, for example, a simple prompt (``a woman in a kitchen'') forces the model to decide the woman’s age, race, body shape, and clothing, the architectural style of the kitchen, its lighting, appliances, and décor ~\citep{Bianchi}. Because the model cannot output an image with unspecified attributes, it defaults to conventions embedded in the training data. \citep{Gillespie2024PoliticsVisibility, yang2025prompts, bhyravajjula-etal-2025-much}. Here, the core worry is the hidden normative work done by the model, and mitigation is about making those dimensions explicit: systems could surface a set of clearly labeled alternatives (different genders, races, body types, family forms, settings), and interfaces could expose controllable attributes so users actively choose how ambiguity is resolved, with options to “diversify” or randomize instead of defaulting to a single normative template \citep{shaw2025agonistic}.

\section{Conclusion}

Ambiguity performs crucial work across domains --- from art and education to law and democratic life ---  by sustaining social friction, richness, plurality and normativity. As LLMs become embedded in interpretive pipelines, a central risk is that they replace this productive openness with decisive, singular resolutions. Our core contribution is to identify and theorize that pattern -- \textit{ambiguity collapse} -- and to specify when it becomes epistemically costly, offering a framework for recognizing an underappreciated set of epistemic harms that arise when models settle meaning. We threaded examples throughout the paper both to show that these risks are already materializing in consequential settings and to convey the stakes involved. We hope this framework supports empirical work that measures these harms in practice and evaluates interventions. The path forward, we argue, is to build sociotechnical arrangements in which models help users navigate contested concepts while preserving the conditions under which those concepts can continue to be argued over, revised, and legitimately stabilized -- moving from systems that impose false clarity and singularity to systems that support the difficult, necessary work of making sense of an irreducibly ambiguous world.

\section{Generative AI Usage Statement}

The authors used ChatGPT (OpenAI; GPT-4 and GPT-5 series) for light editing, and formatting assistance for Table 1. Gemini’s Nano-Banana model was used to color Figure 1. No generative AI tools were used to generate substantive text, arguments, analyses, or original scholarly content. All authors retain full responsibility for the originality, accuracy, and integrity of the manuscript.

\bibliographystyle{ACM-Reference-Format}
\bibliography{references1}

\end{document}